\documentclass[prd,showpacs,floats,floatfix,nofootinbib]{revtex4}

\usepackage{amsbsy}
\usepackage{amssymb}
\usepackage{amsmath}
\input{epsf}
\usepackage{graphicx}
\usepackage{amssymb}
\usepackage{bm}
\def\spose#1{\hbox to 0pt{#1\hss}}

\def\lta{\mathrel{\spose{\lower 3pt\hbox{$\mathchar"218$}}
     \raise 2.0pt\hbox{$\mathchar"13C$}}}
\def\gta{\mathrel{\spose{\lower 3pt\hbox{$\mathchar"218$}}
     \raise 2.0pt\hbox{$\mathchar"13E$}}}

\def\beq{\begin{equation}}
\def\eeq{\end{equation}}
\def\bea{\begin{eqnarray}}
\def\eea{\end{eqnarray}}

\def\n{{\rm n}}

\def\v{{\rm v}}

\def\s{{\rm s}}

\begin{document}
\title{Quantised vortices and mutual friction in relativistic superfluids}

\author{N. Andersson, S. Wells and J.A. Vickers}
\affiliation{Mathematical Sciences and STAG Research Centre, University of Southampton, Southampton 
SO17 1BJ, UK}

\date{\today}

\begin{abstract}
We consider the detailed dynamics of an array of quantised superfluid vortices in the framework of general relativity, as required for quantitative modelling of realistic neutron star cores. Our model builds on the variational approach to relativistic (multi-) fluid dynamics, where the vorticity plays a central role. The description provides a natural extension of, and as it happens a better insight into, existing Newtonian models. In particular, we account for the mutual friction associated with scattering of a second ``normal'' component in the mixture off of the superfluid vortices.  \end{abstract}


\maketitle

\section{Introduction}

In order to understand observational data for astrophysical neutron stars we need to account for the presence of superfluid components in the star's interior. This is true both for phenomena involving short (dynamical) timescales  --- where superfluidity leads to additional modes of oscillation (akin to the second sound seen in laboratory experiments on Helium) --- and long (secular) timescales --- where superfluidity both quenches the nuclear reactions that lead to cooling due to neutrino emission and at the same time opens up new  channels that may lead to fast cooling during specific epochs. The presence of superfluidity has significant impact on the behaviour and evolution of these extreme systems. In the case of dynamics, superfluidity provides an explanation both for the restlessness seen as timing noise and the occasional glitches observed in young radio pulsars \cite{glitch1,glitch2,glitch3,glitch4}. Meanwhile, on evolutionary timescales, X-ray observations of thermal emission from young neutron stars may provide constraints on the superfluid parameters. A celebrated example of this is provided by the youngest known neutron star in the galaxy, situated in the Cassiopeia A supernova remnant \cite{page,shternin}. 

If we want to model realistic scenarios and make maximal use of our understanding of nuclear physics (relating both to the bulk equation of state at supranuclear densities and the relevant superfluid parameters, like pairing gaps and entrainment parameters \cite{chamel}), then we need to carry out our analysis within general relativity. This is well known, yet key aspects of the theory for neutron-star superfluidity have not been developed to the level we require. A particular case in point concerns the so-called mutual friction, a  dissipative channel which is directly related to the presence of quantised vortices in a rotating superfluid \cite{hv1,alpar,mendell,trev}. Estimates suggest that this mechanism may provide the main damping mechanism for various classes of neutron star oscillations and that it is a likely candidate for limiting the growth of modes that are driven unstable by the emission of gravitational waves via the so-called CFS-mechanism \cite{lindblom1,lindblom2,bryn1,bryn2}. The mutual friction is also thought to be the main coupling mechanism associated with pulsar glitches \cite{alpar,mf1,mf2}. 

In this paper we discuss quantised vortices in superfluids and  provide a detailed relativistic formulation for the vortex mutual friction. At the formal level, our analysis has similarities with \cite{langlois}, but we pay more attention to the connection with underlying microphysics, as required in order to prepare the ground for future quantitative models for superfluid neutron-star dynamics. 

\section{Newtonian model}

In order to provide context for our approach to the relativistic vortex problem, we start by exploring the Newtonian case. This provides useful intuition and illustrates the intimate link between the vorticity conservation and the macroscopic fluid dynamics. 

We assume that we are dealing with an array of aligned vortices, which we can average over in order to arrive at a ``hydrodynamical'' model (ignoring for the moment issues related with vortex ``tangles'' and turbulence \cite{vine,chev,schwa,turb}).   The starting point for the discussion is  the  quantised vorticity vector $\kappa^i$, cf. \cite{fluxco,supercon,lagrange}, which represents the macroscopic rotation of the superfluid\footnote{Throughout the paper, we express vectors in terms of their components in a coordinate basis. Hence, we distinguish between contravariant object and covariants ones, although they are related via the spacetime metric; $\kappa_i = g_{ij}\kappa^j$, where the Einstein summation for repeated indices is assumed. Associated with the metric we have the anti-symmetric Levi-Civita tensor $\epsilon_{ijk}$, which is used to express cross products etcetera, in the usual way. A key reason for using this description is that it extends readily to general relativity.}. As the vortices are quantised, we can assign a vortex number density (per unit area) $\mathcal{N}$ to each fluid element. Doing this, it follows that the macroscopically averaged vorticity can be written
\beq
\label{vorticity}
\mathcal{W}^i =\epsilon^{ijk}\nabla_j p_k =\mathcal{N}\kappa^i \ , 
\eeq
where $p_i $ is the canonical momentum of the superfluid \cite{fluxco}, $\kappa^i=\kappa \hat \kappa^i$ where $\hat{\kappa}^i$ is a unit vector in the vortex direction and $\kappa = h/2$ the quantum of circulation (the factor of $1/2$ arises from the fact that we are ultimately interested in neutron stars, where neutrons become superfluid by forming Cooper pairs). Equation \eqref{vorticity} simply states the Onsager-Feynman quantisation condition.

In absence of mechanisms that create or destroy vortices the vortex number density is conserved. This means that we have 
\beq
\partial_t\mathcal{N} +  \nabla_j^\perp \left(\mathcal{N}u_\mathrm{v}^j\right) = 0 \ . 
\label{vorticon}
\eeq
where the derivative acts in the plane orthogonal to the vortex array;
\beq
\nabla^\perp_i = \perp^j_i \nabla_j = \left( \delta^j_i - \hat\kappa^j \hat\kappa_i \right) \nabla_j 
\eeq
Equation \eqref{vorticon} can be seen as the definition of 
 the collective vortex velocity  $u_\mathrm{v}^i$. Taking the time derivative of \eqref{vorticity} we see that
\beq
\partial_t\mathcal{W}^i =-\kappa^i\nabla_j^\perp\left(\mathcal{N}u_\mathrm{v}^j\right) +\mathcal{N}\partial_t\kappa^i.
\eeq
Next we note that, in absence of other forces acting on the vortices, the vector $\kappa^i$ is Lie transported along the flow $u_\mathrm{v}^i$ (although also in the plane perpendicular to the vortices). This means that
\beq
\label{kapeqnnewt}
\partial_t\kappa^i+\perp^i_j\mathcal{L}_{u_\mathrm{v}}\kappa^j = 0,
\eeq
where the Lie derivative is defined by
\beq
\mathcal{L}_{u_\mathrm{v}}\kappa^i = u_\mathrm{v}^j\nabla_j\kappa^i - \kappa^j\nabla_ju_\mathrm{v}^i.
\eeq

Combining these results with the fact that $\nabla_j\mathcal{W}^j=0$, which is obvious from \eqref{vorticity}, we arrive at the vorticity equation
\beq
\label{vort1}
\partial_t\mathcal{W}^i + \epsilon^{ijk}\nabla_j\left(\epsilon_{klm}\mathcal{W}^l u_\mathrm{v}^m\right) =0 \ .
\eeq 
It is worth noting that, this is the standard vorticity equation which we would obtain if the did not have the projections in  \eqref{vorticon} and \eqref{kapeqnnewt}.
Basically, the result shows that the canonical vorticity, $\mathcal{W}^i$, is locally conserved and advected by the $u_\mathrm{v}^i$ flow. 
If we rewrite the vorticity equation \eqref{vort1} as a total outer product, we see that we must have
\beq 
\label{vort2}
\partial_t p^i - \epsilon^{ijk} \epsilon_{klm}u^\mathrm{v}_j \nabla^l p^m = \nabla^i\Psi \ ,
\eeq
where $\Psi$ is an, at this point, unspecified scalar potential. 

Let us now suppose that the vortices do not move with the bulk flow. Letting the latter be represented by $u^i$ we have a velocity difference $v^i = u_\mathrm{v}^i- u^i$. This enables us to use the vorticity definition \eqref{vorticity} to rewrite \eqref{vort2} as 
\beq
\label{vort3} 
n \left[ \partial_t p^i- \epsilon^{ijk} \epsilon_{klm}u_j\nabla^l p^m
- \nabla^i\Psi \right]
=  - \mathcal{N} n \epsilon^{ijk}   \kappa_j v_k = - f_\mathrm{M}^i
\eeq
where $n$ is the number density associated with the superfluid condensate. Written in this form, the right-hand side of the equation provides the Magnus force exerted on the fluid by the vortices. From the construction it is clear that this effect is only present when the vortices and the fluid are not moving together. However, in order for this to be possible we need some other component in the system, interacting with the vortices to balance the force $f^i_\mathrm{M}$  that would act on the vortices. This is, for instance, the case in the celebrated two-fluid model for superfluid Helium, where the two components are the condensate and thermal excitations (phonons and rotons). A similar  model   applies to neutron stars, where superfluid neutrons co-exist with a conglomerate of charged particles (protons, electrons and muons) in the outer core of the star. 
Ignoring the multi-fluid aspects for a moment, we can compare \eqref{vort3} to the standard Euler equations. Thus we see that $\Psi$ accounts for the pressure (and gravity, as required).

\section{Relativistic vorticity conservation}
\label{releuler} 

This results so far summarise the fluid dynamics of a single superfluid condensate which rotates by forming vortices. The argument outlines the strategy we now want to reproduce in the context of general relativity. 

The ``standard'' approach to relativistic fluid dynamics takes as its starting point the stress-energy tensor, usually in the perfect-fluid form;
\beq
T_{ab} = (p+\varepsilon)u_a u_b + p g_{ab} \ ,
\label{stress}\eeq
where $u_a$ is the fluid four-velocity, $p$ is the pressure and $\varepsilon$ the energy density, and uses the requirement that the divergence of $T_{ab}$  must vanish  to obtain the equations of motion. This procedure is straightforward, but extending it to more complex situations is less so. One reason for this is immediately obvious if we consider the double role of $u^a$ in \eqref{stress}. The four-velocity defines the frame of the observer that measures the energy, temperature etcetera. In the standard setting this tends to be taken to be the rest-frame of the fluid, so that $u^a$ describes the fluid flow as well. In a multi-component system the choice of frame is less obvious. This is clearly illustrated by the classic problem of relativistic heat-flow \cite{cesar1,cesar2}.

More complicated settings require a more adaptable approach. Hence, we base our discussion on Carter's convective variational principle \cite{livrev}, which allows for an arbitrary number of interpenetrating fluid components. In the last few years, this strategy has led to progress on problems involving relativistic superfluids \cite{minimal,hawke,2st}, the vexing issue of causality in heat flow \cite{cesar1,cesar2}, elastic systems \cite{lars,carsten}, non-ideal magnetohydrodynamics  \cite{nils} as well providing new insights into dissipative fluid systems \cite{greg}. 

In the case of a single fluid component, the variational model involves a conserved number flux $n^a$, satisfying
\beq
\nabla_a n^a = 0 \ .
\label{cons}\eeq
The momentum conjugate to this flux, $\mu_a$, (obtained via a variation of the relevant energy functional with respect to the flux) satisfies
\beq
\label{euler}
n^a \omega_{ab} = 0 \ ,
\eeq
where the vorticity  $\omega_{ab}$ is defined to be the anti-symmetrised derivative;
\beq
\label{momdef}
\omega_{ab} \equiv 2 \nabla_{[a} \mu_{b]} \ ,
\eeq
and the stress-energy tensor is given by
\beq
T_{ab} = p g_{ab} + n_a \mu_b \ .
\label{newstress}
\eeq

Noting that $n^a = nu^a$, where $n$ is the number density and $\mu_a = \mu u_a$, with $\mu$ the chemical potential, and making use of the (integrated) first law of thermodynamics
\beq
n\mu = p + \varepsilon \ , 
\eeq
it is easy to show that \eqref{newstress} is equivalent to \eqref{stress}. Moreover, it follows that \eqref{cons} and \eqref{euler} imply $\nabla_a T^{ab} = 0$. The two descriptions of the problem are identical, as they have to be.

The variational model contains the same information as the standard approach, but it is more directly linked to the conservation of vorticity. In fact, the definition of the vorticity, equation \eqref{momdef}, implies that its exterior derivative vanishes;
\beq
\nabla_{[a} \omega_{bc]} = 0\ .
\label{symm}\eeq
Whenever the Euler equation \eqref{euler} holds, this leads to the vorticity being conserved along the flow. That is, we have
\beq
\label{vortcon}
\mathcal L_u \omega_{ab} = 0\ .
\eeq
The upshot of this is that the equations of motion \eqref{euler} can be seen as an integrability condition for the vorticity.

Although it is a slight side-issue as far as the present discussion is concerned, it is worth noting that  \eqref{euler} implies that the flow vector $u^a$ is a zero eigenvalue eigenvector for the vorticity tensor $\omega_{ab}$. This means that the vorticity tensor satisfies the degeneracy condition
\beq
\label{degen}
\omega_{a[b} \omega_{cd]} = 0\ ,
\eeq
which in turn implies that  the vorticity tensor must have rank 2. From this we learn that there exists a tangent subspace of eigenvectors $e^a$ which satisfy
\beq
e^a \omega_{ab} = 0\ ,
\eeq
spanned by a unit worldsheet element tangent bivector \cite{stachel}. In his approach to the vorticity problem, Carter focusses on this bivector \cite{Carter1999}. This leads to a elegant description which provides  useful insight into the geometry of the problem. As we will demonstrate elsewhere, this turns out to be useful if one wants to formulate a description of vortex elasticity. However, in the present context --- where our main interest is in the friction which affects the vortex motion --- it is more natural (at least in the first instance) to focus on the corresponding (quantised) vorticity vector.

\section{Quantised vortices}
\label{grnewt}

In order to obtain a relativistic formulation for the vortex mutual friction, let us parallel the Newtonian analysis and work with the quantised vorticity vector, rather than the tensor $\omega_{ab}$. As usual, the vorticity vector is obtained from the vorticity tensor as; 
\beq
\label{vortvec}
\mathcal W^a = {1\over 2}  \epsilon^{abcd} u_b \omega_{cd} \ .
\eeq
Conversely, we have
\beq
\label{vortten}
\omega_{ab} = - \epsilon_{abcd} u^c \mathcal W^d.
\eeq
For the moment, we are assuming that the vortices move with fluid flow, in which case there is a unique four-velocity $u^a$. We will relax this assumption later.

We can see from \eqref{vortvec} that the vorticity vector is orthogonal to the flow, $u_a \mathcal{W}^a = 0$, and also from \eqref{vortten} that the Euler equation \eqref{euler} holds. Next, we use the conservation of vorticity \eqref{vortcon} to find an evolution equation for $\mathcal W^a$;
 \beq
 \mathcal L_u \mathcal W^a +  \mathcal W^a \left(\nabla_b u^b\right)  - u^a \left( \mathcal W^b \dot u_b \right) = 0, 
 \eeq
 where $\dot u^a = u^b\nabla_b u^a$ is the acceleration. This can be written 
 \beq
 h^a_{\ b} \left[  \mathcal L_u \mathcal W^b + \mathcal W^b \left( \nabla_c u^c\right) \right]= 0, 
 \label{vorteq}\eeq
 where the spacetime projection is given by
 \beq
 h^a_{\ b} = \delta^a_b + u^a u_b \ , 
 \eeq
 and we have made use of
 \beq
 \mathcal L_u \epsilon^{abcd} = - \epsilon^{abcd} \left(\nabla_e u^e \right).
 \eeq
 It is straightforward to show that \eqref{vorteq} reduces to \eqref{vort1} in the Newtonian limit.
 
Let us now express the vorticity in terms of a collection of quantised vortex lines. In analogy with the Newtonian analysis, this means that 
 vorticity vector is written 
 \beq
 \mathcal W^a = \mathcal N \kappa^a
 \eeq
We then have
\beq
\mathcal N \kappa^a = {1\over 2} \epsilon^{abcd} u_b \omega_{cd}
\eeq
and 
\beq
\mathcal N u^a = {1\over 2} \epsilon^{abcd} \kappa_b \omega_{cd}
\eeq
From these results, it follows that 
\beq
\perp_a^b \nabla_b (\mathcal N u^a) = 0
\eeq
where the projection $\perp_a^b$ is the same as in the Newtonian case (which is natural, since $\kappa^a$ is a spatial vector in the frame associated with the flow $u^a$).
Making use of these results we see that 
\beq
\mathcal L_u \omega_{ab} = 0 \quad \longrightarrow \quad \kappa^b \nabla_a \left( \mathcal Nu^a \right) + h^b_a\mathcal L_u \kappa^a = 0 
\eeq
leads to 
\beq
\tilde \perp_b^a \mathcal L_u \kappa^b = 0
\eeq
where the combined projection, into the plane orthogonal to both $u^a$ and $\kappa^a$, is
\beq
\tilde \perp_a^b = h_a^c \perp_c^b= \delta_a^b + u_a u^b - \hat\kappa_a \hat \kappa^b
\eeq
In the Newtonian limit, these results lead back to the conservation law \eqref{vorticon} and the equation for the motion of a single vortex \eqref{kapeqnnewt}. (We only need to keep in mind that, in the relativistic formulation the four-velocity $u^a$ relates to the collective vortex three-velocity  $u_\mathrm{v}^i$ in the Newtonian case, as long as the vortices move with the flow).

Let us now ask what would happen if the vortices do not move with the flow, say due to  friction associated with a second fluid component in the mixture. Then we need to make a distinction between the motion of the (array of) vortices $u_\v^a$ and the bulk flow $u^a$. Introducing the relevant velocity difference $v^a$ we can write the  vortex velocity as 
\beq
\label{grvortvel}
u_\v^a = \tilde\gamma\left(u^a +v^a\right) \ ,  \quad u^a v_a = 0 \ ,  \quad  \tilde\gamma = \left( 1 - v^2 \right)^{-1/2} \ . 
\eeq
Vorticity conservation then implies that  
\beq
u_\v^a \omega_{ab} = 0 \ . 
\eeq
Of course, we can always rewrite this as
\beq
\label{greul}
u^a \omega_{ab} = -v^a\omega_{ab} \equiv {1\over n} f^\mathrm{M}_b ,
\eeq
where the right hand side defines the relativistic analogue of the Magnus force. It is easier to see this correspondence if we make use of the  definition for $\omega_{ab}$, \eqref{vortten}. This leads to 
\beq
\label{grmag}
 f^\mathrm{M}_b  =  -  nv^a\omega_{ab} = n\mathcal N \epsilon_{abcd} v^a u_\v^c \kappa^d = n \mathcal N \epsilon_{bad}  \kappa^a v^d,
\eeq
where we have introduced the short-hand notation (relevant for a right-handed coordinate system moving along the flow)
\beq
\epsilon_{abc} = \epsilon_{dabc} u_\v^d. 
\eeq
Hence, from \eqref{greul} and \eqref{grmag}, we arrive at the final equation of motion 
\beq
\label{grsupeqn}
 n^a \omega_{ab} =    f^\mathrm{M}_b \ .
\eeq
The Newtonian limit of this equation of motion leads us back to \eqref{vort3}, as expected. 

The situation we have just described is, of course, somewhat artificial. In order for the argument to make sense, something must prevent the vortices from moving with the bulk flow. The resolution to this is obvious. The description of a real superfluid tends to require two components, and it is the interaction between the vortices and this second component that effects the relative vortex flow.

\section{A two-fluid model with friction}

In order to design a complete model for vortex friction, we need to consider a system of  (at least) two fluids (one of which rotates by forming vortices). 
As a clear, physically motivated, example we will consider the case where we distinguish the flow of matter from that of  heat/entropy, as in He$^4$. 
Let the first  component have particle density $n$ and the second component $s$, and the corresponding fluxes be $n^a = n u^a$ and $s^a = s u_\mathrm{s}^a$. The first component represents the superfluid condensate and the second  could represent thermal excitations in a laboratory system (or a conglomerate of protons and electrons in the case of a neutron star core).  

The matter component is assumed to be conserved;
\beq
\nabla_a n^a = 0 \ , 
\eeq
but the entropy is not (necessarily);
\beq
\nabla_ a s^a = \Gamma_\s \ , 
\eeq
where $\Gamma_\s\ge 0 $ in accordance with the second law of thermodynamics.

This two-fluid system is governed by an energy functional (de facto the Lagrangian) $\Lambda(n,s)$, from which we obtain the momenta that are conjugate to the individual fluxes
\beq
\mu_a = \left. {\partial \Lambda \over \partial n^a } \right|_{s^a} \ , \qquad \mbox{and} \qquad \mu^\s_a = \left. {\partial \Lambda \over \partial s^a } \right|_{n^a}
\eeq 
The variational derivation also provides the stress-energy tensor
\beq
T_{ab} = \Psi g_{ab} + n_a \mu_b + s_a \mu^\s_b 
\label{twotab}
\eeq
where the (generalised) pressure is 
\beq
\Psi = \Lambda - n^a \mu_a + \s^a\mu^\s_a
\eeq
In general, the model allows for nontrivial relations between momenta and fluxes, as in the case of the entrainment effect which plays a central role for neutron star cores \cite{livrev}. However, in the interest of clarity we will not discuss such features here. 

Assuming that the two fluids are coupled by friction \cite{cesar1,cesar2}, we have two coupled equations of motion \cite{greg}
\beq
n^a \omega_{ab} = \mathcal R^\n _b 
\label{neq}\eeq
and
\beq
s^a \omega^\s_{ab} + \mu^\s_b \Gamma_\s = \mathcal R^\s_b\ , 
\label{seq}\eeq
where $\omega_{ab}$ was defined in \eqref{momdef} and 
\beq
\omega^\s_{ab} = 2 \nabla_{[a} \mu^\s_{b]}
\eeq
and $\mu_a^\mathrm{s} = T u^s_a$, where we identify the temperature as $T=-u_s^s \mu^\s_a$ \cite{cesar1,cesar2}. If we assume that there are no external ``forces'' acting on the system then the left-hand sides of \eqref{neq} and \eqref{seq} add up to the divergence of the stress-energy tensor vanishing. Thus, the combined right-hand sides must  cancel, so we have 
\beq
\label{grfbal}
\mathcal R^\n_a + \mathcal R^\s_a = 0. 
\eeq
Moreover, due to $\omega_{ab}$ being antisymmetric, we  see from \eqref{neq} that 
\beq
n^a \mathcal R^\n_a = 0. 
\eeq
We can  also use \eqref{seq} and \eqref{grfbal} to see that
\beq
\label{combine}
\left( s^a \mu^\s_a \right) \Gamma_\s = s^a \mathcal R^\s_a = - s^a \mathcal R^\n_a = - s^a n^b \omega_{ba}. 
\eeq
As the two fluids do not have to move together we can (again) introduce a relative velocity, $w^a$, such that 
\beq
\label{veldiff}
u_\s^a = \gamma \left( u^a + w^a \right), \qquad u^a w_a = 0, \qquad \gamma = \left( 1 - w^2 \right)^{-1/2}.
\eeq
Using this in \eqref{combine}, we find  
\beq
sT \Gamma_\s = s\gamma w^a n^b \omega_{ba} \ge 0 \ . 
\eeq
We  also see that we need
\beq
s^a \mathcal R^\n_a = s \gamma w^a  \mathcal R^\n_a \ge 0, 
\label{2law}\eeq
which can be satisfied by assuming a friction force
\beq
 \mathcal R^\n_a = \alpha w_a \ , \qquad \mbox{with} \qquad \alpha> 0. 
\eeq

At this point, we can make contact with the discussion of quantised vortices from Section~IV. We then see that  \eqref{grsupeqn} suggests that
 we  identify 
\beq
\label{sfforce}
 \mathcal R^\n_a =n \mathcal N \epsilon_{abc}  \kappa^b v^c = f^\mathrm{M}_a
\eeq
This relation is key to our analysis as it relates the friction in the two-fluid system to the vortex dynamics. It is the relativistic analogue of the starting point for the standard discussion of  mutual friction in Newtonian systems \cite{hv1,alpar,mendell,trev} 

\section{Mutual friction}

From a microphysics point of view, one would expect the mutual friction to arise from the scattering of the second component in the system (the heat/thermal excitations in our example) off of the vortex cores \cite{alpar,mendell,trev}. In order to account for this, we introduce yet another relative velocity;
\beq
\label{relvel3}
u_\v^a = \hat\gamma \left( u_\s^a + q^a \right), \qquad u^a q_a = 0, \qquad \hat \gamma = \left( 1 - q^2 \right)^{-1/2}.
\eeq
There are, of course, only two independent relative flows in the problem.
Combining the relative velocities \eqref{grvortvel}, \eqref{veldiff},  and \eqref{relvel3} we see that
\beq
\tilde \gamma = \hat \gamma \gamma
\eeq
and
\beq
q^a = \gamma \left( v^a - w^a\right).
\eeq
Mesoscopically, after averaging over the vortex array, the vortices move under the influence of two forces. The Magnus force is balanced by dissipative scattering off the normal component. This leads to the force balance;
\beq
\alpha q_a = - \mathcal R^\n_a = - \epsilon_{dabc} u_\v^d \kappa^b v^c \ , 
\eeq
assuming that we ignore the inertia of the vortices (which should be insignificant in most realistic situations). We can rewrite this as
\beq
w_a = v_a + {1\over \eta} \epsilon_{abc} \kappa^b v^c,
\eeq
where 
\beq
\eta = \alpha\gamma/\tilde\gamma \ge 0 
\eeq
is the friction coefficient. In fact, is useful to decompose $\kappa^a$ into  components parallel and orthogonal to the flow $u^a$;
\beq
\kappa^a = \kappa_\parallel u^a + \kappa_\perp^a \qquad \mbox{where} \qquad \kappa_\perp^a u_a = 0
\eeq 
in which case we have 
\beq
w_a = v_a + {1\over \eta} \epsilon_{abc} \kappa_\perp^b v^c.
\label{vforce}\eeq

We now have all the information we need to keep track of the vortices as the system evolves. However, in most practical applications it is convenient to eliminate the vortices from the description \cite{hv1}. To do this, we start by rearranging \eqref{vforce} to find an expression for $v^a$ in terms of $w^a$. Then, we can plug the result back into the expression for $\mathcal R^\n_a$. The method we use to rearrange \eqref{vforce} is exactly the same as in the Newtonian problem \cite{mendell,trev}. In the first step, we find that 
\beq
\epsilon^{eafg} u_e \kappa^\perp_f w_a = \eta \left(v^g - w^g\right) +  {1\over \eta} \kappa_\perp^2 \tilde \perp^g_c v^c, 
\eeq
where 
\beq
\tilde \perp^g_c = \delta^g_c - \hat \kappa_\perp^g \hat \kappa^\perp_c, 
\eeq
with $\kappa_\perp^a = \kappa_\perp \hat \kappa_\perp^a $. The second step leads to 
\beq
\epsilon_{bgcd} \epsilon^{eafg} u^b \kappa_\perp^c u_e \kappa^\perp_f w_a = - \kappa_\perp^2 \tilde \perp^c_d w_c 
= - \eta \epsilon_{bgcd} u^b \kappa_\perp^c w^g  - \left(\eta^2 + \kappa_\perp^2 \right) \left(w_d - v_d\right). 
\eeq
and we arrive at the final result;
\beq
v_d = w_d + \left( {1 \over \eta^2 + \kappa_\perp^2} \right) \left[ \eta \epsilon_{bgcd} u^b \kappa_\perp^c w^g -  \kappa_\perp^2 \tilde \perp^c_d w_c    \right].
\label{vexp}\eeq
We now use this in our Magnus force expression \eqref{sfforce} to find 
\beq
f^\mathrm{M}_a =  \mathcal N \tilde \gamma \left[  \left( {\eta^2  \over \eta^2 + \kappa_\perp^2} \right)
\epsilon_{dabc} u^d \kappa_\perp^b w^c + \left( {\eta  \over \eta^2 + \kappa_\perp^2} \right) \kappa_\perp^2 \tilde \perp^c_a w_c 
\right], 
\eeq
which can be used in \eqref{grsupeqn} to close the  system of equations. 
It is worth noting that it follows from the second law \eqref{2law} that $\eta\ge 0$.


We have reach the endpoint of the model development; the equations that need to be solved in order to model a relativistic superfluid with quantised vortices which cause friction as they interact with a normal fluid component. The final equations that need to be solved to obtain the detailed dynamics can be written in different ways, but a practical way to consider the problem involves the overall energy/momentum conservation;
\beq
\nabla_a T^{ab} = 0 
\eeq 
where the stress-energy tensor for the two-fluid system is given by \eqref{twotab}, and an additional Euler equation for the superfluid;
\beq
n^a \omega_{ab} = f^\mathrm{M}_b
\eeq

\section{Concluding remarks}

The rationale for developing a model for quantised superfluid vortices and mutual friction in general relativity is clear. If we want to describe phenomena like radio pulsar glitches, neutron star seismology and other problems involving the presence of a superfluid component in a quantitative fashion then we need to make use of a realistic equation of state for matter. This demands a fully relativistic stellar model and the associated fluid dynamics. One would obviously not expect the local vortex dynamics to be much affected by relativistic gravity. After all, one could always model a fluid element in a local inertial frame. The main issue in writing down the model is consistency. At the end of the day, the model we have designed may not represent a huge leap forwards. Nevertheless, we have taken a necessary ---  not necessarily trivial --- step towards a better understanding of realistic neutron star superfluid dynamics. 

\acknowledgments
NA and JAV acknowledge support from STFC.

\end{document}